\centerline{ A Simple Set of Separable States in a Commutative Simplex. }
\vskip 1cm 
\centerline {  I. D. Ivanovic}
\centerline { Physics Department,Carleton University, }
\centerline { 1125 Colonel By, Ottawa, ON, K1S 5B6, Canada} 
\vskip 2cm
In this note a very crude but simple approximation to the set of separable states in an arbitrary simplex of commutative states is given using the fact that on the lines connecting the maximally mixed state and an arbitrary pure state the positivity of the partial transpose is both necessary and sufficient condition for separability of states. The lower limit to the volume of separable states in a simplex is slightly improved. \par
\vskip 1cm
 0.  The positive partial transpose (PPT)[1] is an extremely useful toll  for detecting separability. In this note it will be used on the lines connecting pure states and the maximally mixed state where PPT is both sufficient and necessary condition for separability. Most of the procedures which will be used are slight modifications of some more general results and well known procedures.
The result, a convex set of separable states is a slightly larger then the previous similar sets.\par 
\vskip 0.5cm 
 1.  To start with, a commuting simplex  over  an orthogonal ray resolution of the identity in $C^n\otimes C^n$, is  
$$S(\{P_m\})=conv (\{ P_m\}_{m=1}^{n^2}) = \{ \sum _m \alpha _mP_m |\quad \sum \alpha _m =1, \alpha _m\geq 0 \}\eqno (1) $$ where 
$$\sum _{m=1}^{n^2}P_m=I \quad , \quad P_mP_n =\delta _{mn}P_m \quad , \quad tr(P_m)=1\quad $$ 

  With respect to separability it is a well known fact that the states on the lines containing the most mixed state and a pure state are easier  to handle , and among the first  examples inspected were   mixtures $$ \rho (\alpha ) = \alpha P + (1-\alpha ) \rho _o $$ 
where $\rho _o = I/n^2 $ and $P$ projects onto the vector 
$$ |\Psi \rangle = N^{-1/2}\sum |i\rangle \otimes |i\rangle .$$ 
These states are  separable for $\alpha \leq 1/(n+1)$.\par
\vskip 0.5cm
A generalization to an arbitrary pure state is given in [2]. A short  description is the following one. Assume that the Schmidt form of the vector onto which the projector $P=|\Psi \rangle \langle \Psi |$ is projecting is $ |\Psi \rangle = \Sigma  \lambda _k |k\rangle \otimes |k \rangle $. Then, in the basis $\{|i\rangle \otimes |j\rangle \}$ assuming $\sum \lambda _k^2 =1 $ the state 
$$ \rho (\alpha ) = \alpha P + (1-\alpha ) \rho _o $$ 
has a non-negative PT when 
$(1-\alpha)/n^2 \geq \alpha \lambda _k \lambda _r $ for all $k,r$.  Denoting $max(\lambda _k \lambda _r)= M$ we obtain the  value of $\alpha _M= 1/(1+n^2M)$.

An explicit separable construction (cf [2]) may be the following one: e.g. take an  (unnormalized) separable vector $|\psi\rangle \otimes |\phi\rangle $ both factors with components $(\sqrt {\lambda _1}, \sqrt {\lambda _2}... \sqrt {\lambda _n})$ in the Schmidt basis of the entangled state. The projector on this state  has matrix elements $P_{ik,mn} \sim \sqrt {\lambda _i\lambda _k \lambda _m \lambda _n} $.  One can "twirl" [3] this state and obtain  the mixture 
$$  \rho _p = {1\over N} \sum _{k=1}^N ((U_1)^k\otimes (U_2)^k)P((U^{\dagger })^k_1\otimes (U^{\dagger }_2)^k) $$
where  $U_1=U_2^*$ are diagonal unitary operators, complex conjugates of each other. The eigen-values of e.g. $U_1$ should look something like 
 $$(1, e^{\imath k\phi }, e^{\imath (k+d)\phi}, e^{\imath (k+3d)\phi}, e^{\imath (k+7d)\phi },... )$$
where $\phi = 2\pi/N$ and N is a large enough prime. This ensures that all differences in phases are different and the  resulting state has the non-zero off-diagonal elements proportional to off-diagonal elements of the entangled state P. Then 

$$ \rho (\alpha _M)={{(\sum \lambda _k )^2}\over {1+n^2M}} \rho _p + \sum _{k\not=r}{{(M-(1-\delta _{kr})\lambda _k \lambda _r)}\over{(1+n^2M)}} |k\rangle \langle k|\otimes |r\rangle \langle r|= $$
$$ = {1 \over {1+n^2M}}P + {{n^2M}\over {1+n^2M}}\rho _o $$

is the first PPT state on the segment $\{P,\rho _o\}$ expressed as a separable state. One should also notice that $\rho _p$ in general, need not belong to the initial simplex and the addition of diagonal elements brings $\rho (\alpha _M)$ back into the initial simplex. On the other hand $\rho _p$ itself gives a decomposition of an entangled state into a sum of separable projectors.  
\vskip 0.5cm
  Extending the segment $\{P,\rho _o\}$  beyond $\rho _o$ one gets the last point which is still a state, $\rho _{p^c} =(I-P)/(n^2-1)$,  the baricenter of the face complementary to P, which is also separable. This state has PPT [4] and   an explicit construction of its separability may be the following one :\par
 first note, (cf. [5]), that the matrix (note the reversed order of $r$ and $k$)
\eject
$$ A_{[kr]} = {1\over { ( \lambda _k+\lambda _r)^2}}( \lambda _r^2 |kk\rangle \langle kk|  -\lambda _k \lambda _r\ (|kr\rangle \langle rk| + |rk\rangle \langle kr| + $$
$$ + \lambda _k \lambda _r (|kr\rangle \langle kr|+|rk\rangle \langle rk|)
+ \lambda _k^2 |rr\rangle \langle rr|) $$
is a separable state in the subspace defined by 
$$(|k\rangle \langle k|+|r\rangle \langle r|)\otimes (|k\rangle \langle k|+|r\rangle \langle r|).$$ 
 Now, the desired state is 
$$ \rho _{p^c} = \sum _{k\langle r} {{( \lambda _k+\lambda _r)^2}\over {n^2-1}} A_{[kr]} + \sum _{k\neq r} {{1-( \lambda _k \lambda _r)}\over {n^2-1}} |kr\rangle \langle kr| .$$
Again, the first term may be outside of the simplex.

  Therefore, in an arbitrary commuting simplex (eq.(1)),the convex hull $$conv (\{(\alpha _mP_m +(1-\alpha _m)\rho _o , (I-P_m)/(n^2-1))_{m=1}^{n^2}\})\eqno (2) $$
is a crude but simple approximation to the set of separable states .\par

 In both constructions one has to "pad" the matrix with diagonal elements, so before the "padding", the states are "farther" away from $\rho _o$ and this simple approximation  could be improved if some of these states belong to the initial simplex. The next improvement may come from the PT of the said simplex, again assuming the PT states belong to the original simplex. \par
  \vskip 0.5cm
3. One can also obtain a slight improvement to the lower limit of the volume of separable states in a arbitrary simplex. The set (2) is made out  of one central simplex, centered at $\rho _o$ and $n^2$ smaller simplices, each having an $(n^2-1)$ order face of the central simplex as a base and the baricenter of the $n^2-1$ base as a vertex. To keep things simple one can make the central simplex a regular one by assuming that the initial simplex is scaled by factor $\alpha _{min}$. The volume of this contraption gives a crude lower limit to the volume of separable states in a simplex as  
$$ V_{sep} \quad  > \quad   {{ \alpha _{min}^{n^2-2} \sqrt {n^2-1}}\over {(n^2-1)!}}\times \big ( \alpha _{min} \sqrt {{n^2}\over {n^2-1}} + n^2 {{(1-\alpha _{min})}\over {\sqrt {n^2(n^2-1)}}}\big )$$
where $\alpha _{min}$ is the minimal $\alpha _k$ from eq. (2), first factor is  the base of the central simplex divided by the dimension, the  first term in the brackets is the height of the central simplex and the second, the sum of heights of $n^2$ outer simplices. \par      
\vskip 0.5cm
   To conclude, the procedures used to obtain set (2)  can be applied to an arbitrary nonorthogonal resolution of the identity, or, as a matter of fact, to an arbitrary set of pure states.
\vfill \eject 
\vskip 0.5cm
 E-mail: igor @ physics.carleton.ca \par
\vskip 0.5cm

 {\bf References:}\par
\vskip 0.5cm
[1] A. Peres: Phys. Rev. Lett. {\bf 77}, 1413 (1996)\par
[2] G. Vidal, R. Tarrach, Phys.Rev.A59 141 (1999)\par
[3] K.G.H. Wolbrecht, R. F. Werner, quant-ph/0010095\par
[4] K. Zyczkowski, P. Horodecki, A. Sanpera, M. Lewenstein,\par
 \hskip 0.5cm Phys. Rev. A {\bf 58}, 883, (1998)\par
[5] A. O. Pittenger, M. H. Rubin, quant-ph/0103038\par
\bye